\documentclass[12pt]{article}
\usepackage{epsfig}

\def\beq{\begin{equation}}
\def\eeq#1{\label{#1}\end{equation}}
\def\eeqn{\end{equation}}

\def\beqa{\begin{eqnarray}}
\def\eeqa#1{\label{#1}\end{eqnarray}}
\def\eeqan{\end{eqnarray}}

\let\bar=\overbar

\def\Dslash{\not{\hbox{\kern-4pt $D$}}}
\def\dslash{\not{\hbox{\kern-2pt $\del$}}}

\def\msb{{\bar{\ssstyle M \kern -1pt S}}}

\def\Title#1{\begin{center} {\Large {\bf #1} } \end{center}}

\begin{document}

\Title{The QCD Phase Diagram and Explosive Astrophysics}

\bigskip\bigskip

%+\addtocontents{toc}{{\it D. Reggiano}}
%+\label{ReggianoStart}

\begin{raggedright}

{\it Stephen D.H. Hsu\index{Hsu}\\
Department of Physics\\
University of Oregon\\
Eugene, OR USA}
\bigskip\bigskip
\end{raggedright}

\section{Introduction}

I was asked by the organizers to give an overview of the QCD phase
diagram, geared towards a mixed audience of astrophysicists and
particle theorists. I chose to emphasize the phase transition(s)
from the normal nuclear phase to the color superconducting and
chirally symmetric phases at high density and low
temperature\footnote{By low temperature here I mean low relative
to the QCD scale, or $\sim 100$ MeV; by astrophysical standards
supernova temperatures of tens of MeV are obviously quite high.}.
I did so because it is possible that these transitions might occur
during violent astrophysical phenomena such as supernovae or
neutron star mergers \cite{HHS}, with potentially dramatic
consequences. If so, these astrophysical events could provide a
window into the physics of QCD.

This contribution is organized as follows. In section 2 I
summarize our current understanding of the QCD phase diagram. In
section 3 I discuss the phase transition characteristics most
relevant to astrophysics, such as critical temperature and latent
heat. In section 4 I discuss supernova core collapse and argue
that conditions at core bounce might lead to the crossing of a QCD
phase boundary. I also briefly discuss neutron star mergers and
hypernovae. I conclude with some discussion from the conference.

\section{QCD phase diagram}

\begin{figure}[htb]
\begin{center}
\epsfig{file=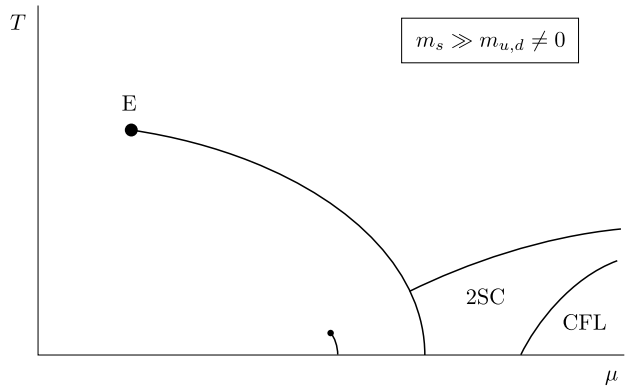,height=2in} \caption{QCD phase diagram
(simplified) in the density-temperature plane.} \label{fig:phase}
\end{center}
\end{figure}

Figure \ref{fig:phase} (taken from Rajagopal and Wilczek's
excellent review in \cite{RW}) shows a possible QCD phase diagram
as a function of temperature and chemical potential. The region at
low temperature and density (lower left) is the normal nuclear
matter phase (the small spur at the bottom is the phase boundary
for the formation of nuclear matter). At sufficiently high density
and low temperature (lower right) it has recently been established
that the ground state of quark matter exhibits Color
Superconductivity (CSC), resulting from the Cooper pairing of
quark quasiparticles near the Fermi surface \cite{RW} --
\cite{Interface}. At high temperature and low density (upper left,
extending to the upper right) we expect to find the quark-gluon
plasma. I will comment on the more specific features of the
diagram below.

%%%%%%%%%%%%%%%%%%%%%%%%%%%%%%%%%%%%%%%%%%%%%%%%%%%%%%%%%%%%%%%%%%%%%%%%%
%%
%%   use this format to include an .eps figure into your paper
%%
%%%%%%%%%%%%%%%%%%%%%%%%%%%%%%%%%%%%%%%%%%%%%%%%%%%%%%%%%%%%%%%%%%%%%%%%%%%

\begin{figure}[htb]
\begin{center}
\epsfig{file=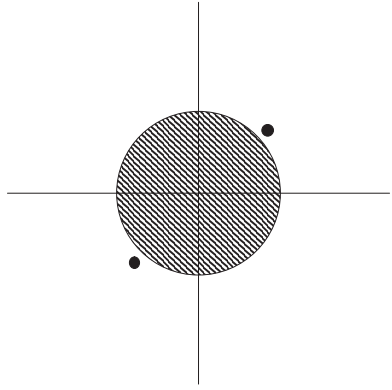,height=2in} \caption{Excitations on opposite
sides of the Fermi surface.} \label{fig:FS}
\end{center}
\end{figure}

Let me try to give you some flavor of the recent results on color
superconductivity. Recall that Cooper pairing (as in ordinary
superconductors) involves excitations near the Fermi surface, but
with equal and opposite momenta (figure \ref{fig:FS}). In QCD the
excitations have quark quantum numbers, and hence there is a
directly attractive channel due to gluon exchange in the
anti-triplet ($\bar{3}$) color representation. In the case of
electrons in a metal photon exchange is repulsive, and the
attractive channel is due to the exchange of phonons (lattice
vibrations).

What is common between quark matter and an ordinary metal is the
existence of a Fermi surface. The Pauli exclusion principle
requires that fermions like quarks or electrons occupy distinct
quantum states, hence at zero temperature the ground state
consists of filled levels up to some energy (this boundary is the
Fermi surface). The lowest energy excitations of this system,
called quasiparticles, are states just above the Fermi surface.
Now, a Fermi surface is unstable with respect to attractive
interactions in the Cooper pairing channel. That is, even an
arbitrarily weak interaction in this channel can lead to pairing
of quasiparticles, which reduces the overall energy of the system.
Heuristically, the spherical symmetry of the Fermi surface (in
momentum space, the energy of a quasiparticle only depends on its
distance from the surface, and not on its angular position)
reduces the pairing dynamics to that of a 1+1 dimensional system,
where even weak interactions lead to pairing.

%%%%%%%%%%%%%%%%%%%%%%%%%%%%%%%%%%%%%%%%%%%%%%%%%%%%%%%%%%%%%%%%%%%%%%%%%
%%
%%   use this format to include an .eps figure into your paper
%%
%%%%%%%%%%%%%%%%%%%%%%%%%%%%%%%%%%%%%%%%%%%%%%%%%%%%%%%%%%%%%%%%%%%%%%%%%%%

Now, recall that QCD exhibits a property called asymptotic
freedom, which means that short distance (large momentum transfer)
interactions are weak, whereas long distance (small momentum
transfer) interactions are strong. At very high quark densities,
most interactions occur over short distances, and we therefore
have very good control over calculations. However, at quark
densities likely to be found in a neutron star, roughly a few per
cubic Fermi, the typical interactions are quite strong, precluding
reliable quantitative results. Nevertheless, the indications of at
least an {\it instability} to the formation of Cooper pairs is
still evident, even if we can't say more about the details.

\begin{figure}[htb]
\begin{center}
\epsfig{file=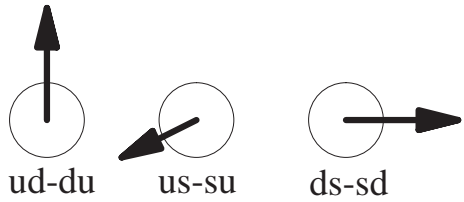,height=1in} \caption{Di-quark orientation in
the CFL phase. The arrow represents a $\bar{3}$ orientation in
color space.} \label{fig:N3}
\end{center}
\end{figure}

In figure \ref{fig:phase} there are two distinct CSC phases
displayed, the 2SC and CFL phases. In the former, the up and down
quarks pair in the isosinglet channel, and the strange quark is
left to pair with itself in an exotic (possibly spin 1) channel.
In the latter, the flavor and color orientations of all three
quarks are correlated in a non-trivial way (see figure
\ref{fig:N3}): pairing in a particular flavor channel corresponds
to a particular orientation in color space. In this talk I will
assume that figure \ref{fig:phase} is correct in that the 2SC
phase has the lowest energy at intermediate density. The
transition from nuclear matter to the 2SC phase is likely to be
first order \cite{Universal}, as we will discuss further below.

%%%%%%%%%%%%%%%%%%%%%%%%%%%%%%%%%%%%%%%%%%%%%%%%%%%%%%%%%%%%%%%%%%%%%%%%%
%%
%%   use this format to include an .eps figure into your paper
%%
%%%%%%%%%%%%%%%%%%%%%%%%%%%%%%%%%%%%%%%%%%%%%%%%%%%%%%%%%%%%%%%%%%%%%%%%

Before leaving the subject of the QCD phase diagram, let me
discuss another transition -- the chiral phase transition -- that
is likely to occur as we increase the baryon number density.
Chiral symmetry has to do with unitary rotations among different
flavors of quarks:
\begin{equation}
q_i^{L,R} \rightarrow U_{ij} q_j^{L,R}~~~,
\end{equation}
where i labels the flavor (i = up, down, strange) and $U$ is a
unitary matrix. L and R label whether the quark is left handed or
right handed. Chiral symmetry reflects the fact that, in the limit
of small quark masses, the strong interactions (gluons) can't tell
the difference between different flavors of left and right handed
quarks.

These symmetries are broken by quark condensation in the QCD
ground state:
$$
\langle \bar{q}_L q_R \rangle \neq 0~~~,
$$
but are restored in the 2SC phase and in the quark-gluon plasma.
(However, in the CFL state chiral symmetries do remain broken at
high density.) Their restoration involves a phase transition,
again at quark densities of roughly a few per cubic Fermi. What
can we say about this transition?

Because it occurs at a density where the effective coupling is
strong, we can't make quantitative predictions. However, we can
learn something about the order of the transition from something
called ``universality''. The logic is as follows: if the
transition is second order (or higher), then at the precise point
of transition there must be very long wavelength excitations in
the system. A heuristic way to understand this is to visualize the
effective potential near the transition. In a second order
transition the potential becomes very flat as the second
derivative changes sign from positive to negative. (In contrast,
in a first order transition the system tunnels or fluctuates out
of the disfavored vacuum while the second derivative is still
positive.)

Technically, a second order transition implies fluctuations of
infinite wavelength (or equivalently, zero mass excitations).
There are not very many models which describe such low-energy
dynamics. To be self-consistent they need to exhibit something
called an infrared fixed point in the renormalization group
evolution of the coupling constant. Roughly speaking, the
low-energy dynamics must become almost scale invariant as the
correlation length diverges. The candidate models describing the
low-energy dynamics must possess the same symmetry properties as
the underlying system, but otherwise can be quite simple. In some
cases, there are no candidate models with the right symmetries and
low-energy particle content that exhibit an infrared fixed point.
In this case the transition is predicted to be first order. Based
on this kind of analysis, both the 2SC and chiral phase transition
at finite density are likely to be first order
\cite{ChPT,Universal}.

So, we have two candidate phase transitions which, as we discuss
below, might have astrophysical implications. Obviously, a number
of issues are still unresolved. For example, are there two
separate phase transitions, or does chiral symmetry restoration
coincide with color superconductivity? The renormalization group
analysis of color superconductivity \cite{ESH,SW} suggests that
once the effective excitations start to look like quarks, there is
a pairing instability and the system is likely to be
superconducting. However, one could also imagine a chirally
restored phase with nucleonic excitations, so chiral symmetry does
not necessarily imply CSC. In any case, both transitions are
likely to occur at temperatures and densities that might be
achieved by astrophysical phenomena.

\section{Phase transition parameters}

At asymptotic density, the nature of the CSC order parameter, the
binding energy density and the critical temperature are all
precisely calculable. Less is known about intermediate densities
of several times nuclear density. However, there are strong
indications of a Cooper pairing instability, and estimates of the
resulting gap are of order $\Delta \sim$ 40-120 MeV at a quark
chemical potential $\mu$ of $\approx$ 400 MeV. The nature of the
chiral phase transition is similarly poorly determined, although
as we discussed in the previous section it is likely to be first
order. Rather than discuss the two transitions separately, we will
focus on the CSC transition, keeping in mind that if a first order
chiral boundary is crossed, the consequences should be similar.

Consider the possible implications of CSC for the collapse and
explosion of massive stars ($M > 8 M_{\odot}$). We will argue that
it is quite likely that at the moment of maximum compression of
the collapsing Fe core, the densest part of the star crosses the
critical density into the phase where CSC is energetically
favored. The release of latent energy has the potential to
generate an explosive shockwave which powers the resulting
supernova (SN). Current simulations of supernovae are generally
unable to reproduce the explosive behavior observed in nature: the
shockwave generated by the mechanical bounce of the nuclear core
stalls before reaching the surface, unless an appeal is made to
neutrino reheating combined with non-spherical phenomena such as
rotation or convection \cite{SNcollapse,SNsim}. We also note that
the energy liberated in a CSC phase transition is potentially
sufficient to power hypernovae (HN) \cite{Hyper}, which have been
linked to gamma ray burst events \cite{GRB}.

First, let us summarize some results on CSC from the recent
literature. Precise results are only valid at asymptotic densities
where the effective QCD coupling is small, however they should
still be useful guide when dealing with intermediate densities. In
any case, our qualitative results will be insensitive to factors
of 2 in these formulas:

\bigskip

\noindent $\bullet$ Gap size: $\Delta \sim$ 40 - 120 MeV

\smallskip

\noindent $\bullet$ Critical temperature: $T_c \simeq .57 \Delta$

\smallskip

\noindent $\bullet$ At asymptotic density the binding energy
density is $E_{CSC} = {\frac{5.8}{4 \pi^2}} \Delta^2 \mu^2$.
Simple dimensional analysis (given the absence of any small
parameter) also suggests a value in this range. Note that we are
interested here in the latent {\it energy} associated with the
first order transition to the CSC phase (or any other transition
that occurs as the baryon density is increased beyond several
times nuclear density). The baryon density changes on
astrophysical timescales, or very slowly on the timescale of QCD
dynamics, and the vacuum state at zero temperature (or at $T <<
T_c$) is found by minimizing the energy in the sector of the
Hilbert space with fixed baryon number. Studies such as
\cite{Interface}, which involve the free energy $\Omega$ at finite
chemical potential (but not fixed density) are appropriate for
determining pressure equilibrium between nuclear and CSC matter in
circumstances in which baryon number can flow across a boundary
(e.g. in a neutron star), but do not address a possible SN
transition. (See discussion for further remarks.)

\smallskip

\noindent $\bullet$ Phase diagram: the normal nuclear phase is
separated from the CSC phase (most likely the 2SC two flavor
condensate phase, although possibly the CFL phase
\cite{ChPT,Universal,Interface}) by a first order boundary.

\bigskip

\section{Astrophysics of Core Collapse}

Now let us review the standard scenario of Fe core collapse which
is believed to lead to type II supernovae \cite{SNcollapse}.
Nuclear burning during the $10^7$ year lifetime of the star leads
to a shell structure, with the inner core eventually consisting of
Fe ash. Because iron cannot participate in further exothermic
nuclear reactions, there is an eventual cooling and collapse of
the Fe core, whose mass is likely to be $(1-2)~ M_{\odot}$ (or,
roughly the Chandresekhar mass). The collapse of this core is only
halted by neutron degeneracy, which leads to a stiffening of the
equation of state. The resulting bounce produces a shockwave,
whose energy of $\sim 10^{51}$ ergs is a small fraction of the
total available gravitational binding energy released by the
collapse:
\begin{equation}
E_b \sim 3 \cdot 10^{53} {\rm ergs} \left( {\frac{M_{core}
}{M_{\odot}}} \right)^2 \left( {\frac{R }{10 {\rm km}}}
\right)^{-1} \ . \end{equation} Most of this energy escapes in the
form of neutrinos during the supernova, as was observed in the
case of SN1987a.

The pressure in the collapsed core at the instant of the bounce is
most likely {\it greater} than the corresponding pressure in any
remnant neutron star. In order to cause a bounce, the kinetic
energy $E_b$ of the infalling material (which is a sizeable
fraction of a solar mass!) must be momentarily stored as
compressional potential energy in the (sub)nuclear matter. This
additional mechanical squeezing at the bounce suggests that if the
critical density for CSC is ever reached in a neutron star, it
will be reached at this instant.

Simulations of the core bounce result in densities of at least
several times nuclear density ($5-10 \cdot 10^{14} {\rm g/cm^3}$),
and temperatures of roughly 10-20 MeV \cite{SNcollapse}. This
temperature is likely less than $T_c$ for CSC, possibly much
less\footnote{It is conceptually easier to think about the case
where $T$ is much less than $T_c$, since in this case the Free
energy ($F = U - TS$) liberated by the transition is predominantly
energy, with only a small component related to entropy. The
relevant dynamics is governed by energetics rather than Free
energetics.}, and hence the core of the star traverses the phase
diagram horizontally in the density-temperature plane, crossing
the critical density boundary into the CSC phase. It is important
to note that the core region at bounce is probably {\it cooler}
than post-bounce, since degenerate neutrinos tend to heat the
proto-neutron star as they diffuse out \cite{Lattimer}. Studies
quoting larger SN temperatures such as $T \sim 30$ MeV generally
refer to this later stage \cite{Carter}.

Once the core crosses into the CSC part of the phase diagram, the
transition proceeds rapidly, on hadronic timescales. Because the
transition is first order, it proceeds by nucleation of bubbles of
the CSC phase in the normal nuclear background. The rate of bubble
nucleation will be of order $({\rm fm})^4$ (${\rm
fm}=10^{-13}$cm.), due to strong coupling. (In a system governed
by a dimensional scale $\Lambda$, the nucleation rate is given by
$\Gamma \sim \Lambda^4 e^{-S}$, where $S$ is the action of the
Euclidean bounce solution interpolating between the false (normal
nuclear) vacuum and a bubble of critical size. At strong coupling,
$S$ is of order one, so there is no large exponential suppression
of the nucleation rate. The scale $\Lambda$ is of the order of
$\Delta$ or $\mu = 400$ MeV.)

Causality requires that the mechanical bounce of the core happen
over timescales larger than the light crossing time of the core,
or at least $10^{-4} s$. Hence, the phase transition occurs
instantaneously on astrophysical timescales. A nucleated bubble of
CSC phase expands relativistically -- liberated latent heat is
converted into its kinetic energy -- until it collides with other
bubbles. Because the system is strongly coupled, these collisions
lead to the rapid production of all of the low energy excitations
in the CSC phase, including (pseudo)Goldstone bosons and other
hadrons. The resulting release of energy resembles an explosion of
hadronic matter.

To estimate the total CSC energy released in the bounce, we use
the result that the ratio of CSC binding energy density to quark
energy density is of order $\left(\frac{ \Delta }{\mu }\right)^2$.
For $\mu \sim 400$ MeV, and $\Delta \sim$ 40-120 MeV, this ratio
is between .01 and .08, or probably a few percent.
\begin{equation}
E \sim \left(\frac{ \Delta }{\mu }\right)^2 M_{core} \ .
\end{equation}
In other words, the total energy release could be a few percent of
a solar mass, or $10^{52}$ ergs! This is significantly larger than
the energy usually attributed to the core shockwave, and possibly
of the order of the gravitational collapse energy $E_b$. The
implications for SN simulations are obviously quite intriguing.

In \cite{BH} it was suggested that strange matter formation might
overcome the energetic difficulties in producing type-II supernova
explosions. While there are strong arguments that the CSC
transition should be first order, and reasonable order of
magnitude estimates of the latent heat \cite{RW,Interface}, it is
not clear to us why there would be supercooling in a strange
matter transition. The conversion of up quarks to strange quarks
must proceed by the weak interactions, but the rate is still much
faster than any astrophysical timescale. Thus, the population of
strange quarks is likely to track its chemical equilibrium value
as the pressure of the core increases. There may be an important
effect on the nuclear equation of state from strangeness (e.g. a
softening of the pressure-density relationship), but we do not see
why there should be explosive behavior.

Our results are also relevant to hypernovae \cite{Hyper}, which
are observed to have ejecta kinetic energies 10-100 times larger
(of order $10^{52-53}$ ergs) than those of ordinary type II
supernovae. Accounting for this extra kinetic is extremely
challenging in standard scenarios. However, for exceptionally
massive stars with $M < 35  M_{\odot}$ (for $M > 35 M_{\odot}$ the
hydrogen envelope is lost during H-shell burning, and the core
size actually decreases~\cite{chlee}) there is a large core mass
which leads to a larger release of CSC binding energy. In fact,
the released energy might depend nonlinearly and sensitively on
the star's mass at the upper range. For example, the fraction of
$M_{core}$ which achieves critical density might be a sensitive
function of the mass of the star.

Another alternative is that hypernovae are the result of neutron
star mergers rather than the explosion of an individual star. This
possibility has been examined in relation to hypernovae as the
engines of gamma ray bursts (GRBs) \cite{GRB}. It seems quite
plausible that in the merger of two cold neutron stars a
significant fraction of the stars' mass undergoes the CSC
transition (i.e. crosses the critical pressure boundary for the
first time; in this case the temperature is probably negligible
relative to the CSC scale $\Delta$). This provides a substantial
new source of energy beyond gravitational binding, and may solve
the ``energy crisis'' problem for this model of GRBs \cite{GRB}.

Finally, we note that the trajectory of the SN core in the
temperature-density phase diagram might be rather complicated. The
parameters suggest a density transition (at $T < T_c$), but
subsequent reheating of the core due to the explosion, or to
neutrino diffusion \cite{Lattimer} might raise the temperature
above $T_c$, and lead to additional transitions across the
temperature boundary \cite{Carter}. When $T \sim T_c$, the Free
energies ($F = U - TS$) of the normal and CSC phases are
comparable, due to the larger entropy of the normal phase. The
evolution of bubbles in this regime is governed by relative Free
energies rather than energetics alone, and the transition is
presumably less explosive than the pressure transition at $T <<
T_c$.

\section{Summary}

Our understanding of QCD at high density has evolved dramatically
over the past two years, leading to remarkable progress in
understanding the QCD phase diagram and the color superconducting
state of quark matter. We have argued here that phase transitions
at high density and low temperature (relative to the QCD scale)
might play a role in violent astrophysical phenomena. In
particular, the densities and temperatures associated with core
collapse in massive stellar evolution suggest that CSC could play
an important role in type II supernovae, or possibly hypernovae
(GRBs).

Our assumptions concerning key parameters are conservative, and
taken from distinct (and heretofore independent) regimes of
inquiry: stellar astrophysics and dense quark matter. Yet, they
point to the interesting possibility that supernova explosions are
powered by CSC binding energy. It is well established that the
shockwave energy from core collapse is insufficient to produce an
explosion, and recent results incorporating Boltzman transport of
neutrinos show that neutrino reheating is also insufficient unless
non-spherical phenomena such as rotation or convection are taken
into account \cite{SNsim}. We are optimistic that future progress
in simulations will tell us much about whether and how latent
energy from QCD plays a role in stellar explosions.

\bigskip \noindent {\bf Acknowledgements}

\noindent I would like to thank F. Sannino and R. Ouyed for
organizing this very stimulating workshop. The work of S.H. was
supported in part under DOE contract DE-FG06-85ER40224 and by the
NSF through the Korean-USA Cooperative Science Program, 9982164.

%\newpage

\newpage

\def\Discussion{
\setlength{\parskip}{0.3cm}\setlength{\parindent}{0.0cm}
     \bigskip\bigskip      {\Large {\bf Discussion}} \bigskip}
\def\speaker#1{{\bf #1:}\ }
\def\endDiscussion{}

\Discussion

\speaker{Student} How does one identify the order of a phase
transition? What is the difference between first and second order?

\speaker{Hsu} Formally speaking, the order refers to the level of
discontinuity exhibited by thermodynamic quantities (in the
infinite volume limit) at the transition. In an Nth order
transition, the Nth derivative of the free energy (with respect
to, e.g., temperature or density) is discontinuous, while lower
order derivatives are continuous. So, in a first order transition
the derivative of the free energy itself is discontinuous, whereas
in a second order transition the first derivative of the free
energy is continuous and the second derivative is not. One can
imagine even smoother transitions where only some high derivative
of the free energy is discontinuous. Alternatively, one can
distinguish by noting that the order parameter exhibits a jump in
the first order case, but evolves smoothly in the 2nd and higher
order cases. Of course, it is only first order transitions (with
nonzero latent heat) that can lead to the explosive phenomena
considered here.

\bigskip

\speaker{Mark Alford (University of Glasgow)}  (paraphrasing) In
our recent investigations of a model of the nuclear matter--quark
matter interface in neutron stars, we find that while the free
energy of the quark phase is lower, the energy density of the
quark core is actually higher than in the nuclear phase. If this
is the case, how does the phase transition proceed?

\speaker{Hsu} In this discussion I assumed that there is a high
density state (either the CSC or chirally restored phase) with
lower energy than the normal nuclear phase. Note that I really
mean energy here, not free energy: $F = H - \mu N = - P$. If we
consider the phase diagram as a function of baryon density (not
baryon chemical potential), the existence of a new phase implies a
state with lower energy than the normal nuclear phase at the same
density. If the transition is first order, there is the
possibility of latent energy release. In a core bounce, the baryon
density is increased on a timescale too short for bulk baryon
number flow. So, the relevant question is what happens when you
cross a boundary in the baryon density direction. The system will
try to tunnel into the lower energy phase, and the release of
latent energy powers the explosion.

On the other hand, if the compression is slow enough for
significant rearrangement of baryon number, then it is governed by
free energetics, or $F$. In the model you considered, the quark
matter phase has lower free energy, but higher energy and baryon
density than the nuclear phase. The system can only form the quark
phase by aggregating baryon number, and there is probably not
enough time for this to happen during a supernova core bounce. The
conversion from nuclear to quark phases also requires energy
input, which presumably has to come from gravitational collapse.

\endDiscussion

\end{document}